\DeclareUnicodeCharacter{2061}{}
\documentclass[journal]{IEEEtran}
\usepackage{amsmath,amsfonts,bm}
\usepackage{algorithm}
\usepackage{algorithmic}
\usepackage{CJKutf8}
\usepackage{array}
\usepackage{footnote}
\usepackage{textcomp}
\usepackage{stfloats}
\usepackage{url}
\usepackage{verbatim}
\usepackage{graphicx}
\usepackage{cite}
\usepackage{color}
\usepackage[caption=false,font=normalsize,labelfont=sf,textfont=sf]{subfig}
\begin{document}

\title{Reconfigurable Holographic Surface-aided Distributed MIMO Radar Systems}

\author{Qian Li,~\IEEEmembership{Graduate Student Member,~IEEE,} Ziang Yang,~\IEEEmembership{Graduate Student Member,~IEEE,} \\Dou Li, Hongliang Zhang,~\IEEEmembership{Member,~IEEE}

\thanks{Manuscript received 12 November 2024; revised 4 March 2025 and 3 April 2025; accepted 4 April 2025. This work was supported in part by the National Science Foundation under Grant 62371011 and in part by the Beijing Natural Science Foundation under Grant L243002.

The authors are with the State Key Laboratory of Photonics and Communications, the School of Electronics, Peking University, Beijing, China (e-mail: qianli@stu.pku.edu.cn; \{yangziang,lidou,hongliang.zhang\}@pku.edu.cn).}}

\maketitle

\begin{abstract}
Distributed phased Multiple-Input Multiple-Output (phased-MIMO) radar systems have attracted wide attention in target detection and tracking. However, phase-shifting circuits in phased subarrays lead to high power consumption and hardware cost. The reconfigurable holographic surface (RHS) offers an energy-efficient and cost-effective alternative to address this issue. In this letter, we propose RHS-aided distributed MIMO radar systems that provide more accurate multi-target detection with equivalent power consumption and hardware cost compared to distributed phased-MIMO radar systems. The RHS realizes beam steering by regulating the radiation amplitude of its elements, making conventional beamforming schemes designed for phased arrays inapplicable. To maximize detection accuracy, we design an amplitude-controlled beamforming scheme for multiple RHS subarrays. Simulations validate the superiority of the proposed scheme over the distributed phased-MIMO radar scheme and reveal the optimal allocation between coherent processing gain and diversity gain for optimal system performance with fixed hardware resources.
\end{abstract}


\section{Introduction}
\IEEEPARstart{D}{istributed} multiple-input multiple-output (MIMO) radar\cite{16} consists of widely separated antennas that observe targets from different angles, reducing the scintillation of the radar cross section (RCS). By transmitting orthogonal waveforms, it separates all transceiver paths and achieves spatial diversity gain, thereby enhancing target detection performance. However, this comes at the cost of losing the coherence array-processing gain of traditional phased array radar systems. To combine these two gains, the distributed phased-MIMO radar\cite{15} is proposed, which replaces each antenna in the distributed MIMO radar with a phased subarray. However, due to the power-consuming and high-cost hardware circuits composed of phase shifters and power amplifiers in distributed phased-MIMO radars, the system aperture is limited, making it difficult to improve performance.

To address these issues, a new type of metasurface antenna array\cite{25,26} called reconfigurable holographic surface (RHS)\cite{deng2022hdma,27} is developed. The RHS contains numerous metamaterial elements and the beam is controlled by adjusting the radiation amplitude of each element. Specifically, the RHS regulates the beam with simple diodes, which are much more energy-efficient and cost-effective than phase shifters and power amplifiers in phased arrays\cite{13}. Therefore, the RHS can achieve a larger array aperture under the same power consumption and hardware cost as the phased array, thereby enhancing detection accuracy\cite{30}. 

In literature, there are some works that investigate RHS radar systems. In \cite{7}, the authors proposed an RHS-enabled holographic radar and derived an optimal closed-form RHS beamformer with the highest signal-to-noise ratio (SNR). In \cite{9}, the authors proposed an RHS-based radar system and designed a joint waveform and amplitude optimization algorithm for multi-target detection. Besides, several studies\cite{21, 23} investigate the impact of hardware impairments in RHS beamforming design. However, existing studies on RHS radar systems focus on the single-static radar system, and there is a lack of research on RHS-aided distributed MIMO radar systems, which need to jointly design the beamformer of multiple subsystems, making the design of such a scheme more complicated.

In this letter, we investigate RHS-aided distributed MIMO radar systems for multi-target detection. The considered beamforming scheme design presents the following challenges. First, the multi-target/clutter multi-subsystem architecture increases the complexity of system performance evaluation. Second, the RHS achieves beamforming through regulating the radiation amplitude of its elements, making conventional phase-shifting schemes inapplicable. To address these challenges, we first formulate an optimization problem aiming to maximize multi-target detection capability while balancing the performance of all subsystems. We then develop an amplitude optimization algorithm that jointly optimizes beamformers of all RHS subarrays. Simulation results demonstrate the advantages of the proposed scheme compared to distributed phased-MIMO radar systems under equal power consumption
and hardware cost. Moreover, to enhance performance, we simulate the optimal allocation between the spatial diversity gain and the coherent processing gain with fixed hardware resources.

\section{System Model}
This section introduces the proposed system model, the RHS model and the received signal model.

\begin{figure}[!t]
\centering
\includegraphics[width=2.5in]{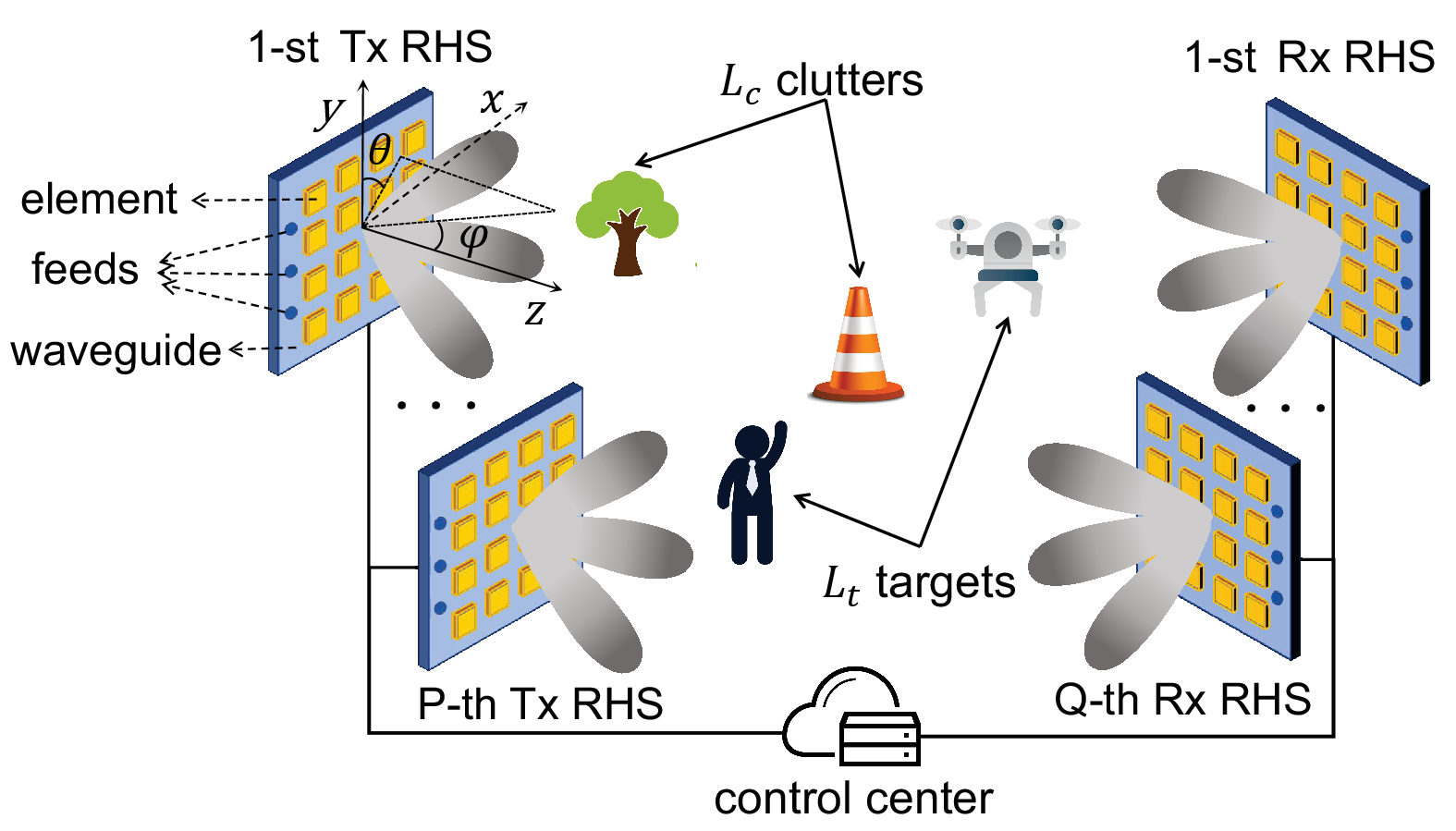}
\captionsetup{font={small,stretch=1.25},justification=raggedright}
\caption{System model for RHS-aided distributed MIMO radar.}
\label{fig_1}
\end{figure}

\subsection{Scenario Description}
As shown in Fig. 1, RHS-aided distributed MIMO radar systems consist of $P$ Tx RHS subarrays, $Q$ Rx RHS subarrays, $L_t$ static targets, $L_c$ clutters and a control center. Each Tx/Rx RHS contains $N_t=N_r=N_{x} \times N_{y}$ elements and $K_t=K_r=K$ feeds. All subarrays are randomly distributed in space. The total number of elements, i.e. hardware resources of the system\cite{15}, is $N_{sum}=P{N_t}+Q{N_r}$. The reflection coefficient of each target/clutter is inhomogeneous, and each target/clutter is in a different direction from each subarray. The control center connects to the subarrays, processing the received signals and controlling the beams of the subarrays. 


\subsection{RHS model}
The RHS is a type of leaky wave antenna that consists of $K$ feeds, a waveguide, and densely arranged sub-wavelength metamaterial elements. The feeds, attached to the edge of the RHS, convert injected signals into electromagnetic waves. The radiation amplitude of each element can be adjusted within the range of [0,1] by changing the bias voltage applied to each element\cite{deng2022hdma}. The RHS beam is the superposition of leaky waves from each element. Therefore, by regulating the amplitude of each element, RHS can flexibly control the beam and achieve holographic beamforming.

Assume that $P$ Tx RHS subarrays respectively transmit $P$ mutually orthogonal narrowband signal waveforms. In the $p$-th Tx RHS, it is assumed that the same signal $\boldsymbol{s}_p\in\mathbb{C}^{1\times I_t}$ is injected into all $K_t$ feeds, where $I_t$ denotes the number of snapshots. Thus the signal $\boldsymbol{S}_p\in\mathbb{C}^{K_t\times I_t}$ injected into the $p$-th Tx RHS consists of a stack of $K_t$ vectors $\boldsymbol{s}_p$. The signal $\boldsymbol{X}_p\in\mathbb{C}^{N_t\times I_t}$ radiated by the $p$-th Tx RHS is as follows: \begin{equation}\boldsymbol{X}_p=\boldsymbol{\mathit{\Psi}}_p^t\left(\boldsymbol{Q}_p^t\circ\boldsymbol{\mathit\Gamma}_p^t\right)\boldsymbol{S}_p,\end{equation} 
where $\boldsymbol{\mathit{\Psi}}_p^t=diag\left\{\boldsymbol{\psi}_p^t\right\}$, with $\boldsymbol{\psi}_p^t=\left[\psi_{p,1}^t,...,\psi_{p,N_t}^t\right]$ as the beamformer of the $p$-th Tx RHS, which is realized by a simple diode circuit controlling the radiation amplitude of each element\cite{27}. The matrices $\boldsymbol{Q}_p^t\in\mathbb{C}^{N_t\times K_t}$ and $\boldsymbol{\Gamma}_p^t\in\mathbb{C}^{N_t\times K_t}$ respectively indicate the inherent phase shift and amplitude attenuation when the reference signal propagates in the waveguide. The symbol $\circ$ is the Hadamard product. The $(n_t, k_t)$-th element of  $\boldsymbol{Q}_p^t$ is $q_{n_t,k_t}^t=e^{-j2\pi\nu D_{n_t,k_t}/\lambda}$, where $D_{n_t,k_t}$ is the distance between the $k_t$-th feed and the $n_t$-th element, $\lambda$ is the wavelength of the signal, and $\nu$ is the refractive index. The $(n_t, k_t)$-th element of  $\boldsymbol{\Gamma}_p^t$ is $\gamma_{n_t,k_t}^t=e^{-aD_{n_t,k_t}}$, where $a$ is the amplitude attenuation factor.

Due to the reciprocity of antennas, the signal received by the $q$-th Rx RHS subarray can be expressed as:
\begin{equation}
\boldsymbol{Y}_q=\left[\boldsymbol{\mathit{\Psi}}_q^r\left(\boldsymbol{Q}_q^r\circ\boldsymbol{\Gamma}_q^r\right)\right]^T\boldsymbol{V},
\end{equation}
where $\boldsymbol{V}\in\mathbb{C}^{N_r\times I_r}$ denotes the echo signal with $I_r$ being the number of snapshots, the matrix $\boldsymbol{\mathit{\Psi}}_q^r=diag\left\{\boldsymbol{\psi}_q^r\right\}$, with $\boldsymbol{\psi}_q^r=\left[\psi_{q,1}^r,...,\psi_{q,N_r}^r\right]$ as the beamforming vector of the $q$-th Rx RHS subarray. Besides, $\boldsymbol{Q}_q^r\in\mathbb{C}^{N_r\times K_r}$ and $\boldsymbol{\Gamma}_q^r\in\mathbb{C}^{N_r\times K_r}$ denote the inherent phase shift and amplitude attenuation of the Rx RHS, respectively, as defined for the Tx RHS.

\subsection{Received Signal Model}
Suppose that $P$ Tx RHS subarrays radiate signals, which propagate to $L=L_t+L_c$ targets/clutters and reflect to $Q$ Rx RHS subarrays. The signal at each Rx RHS is filtered through a matched filter bank $MF_{s_p},p=1,2,...,P$ where $MF_{s_p}=\boldsymbol{S}_p^H$ denotes the matched filter related to the $p$-th transmit signal\cite{16}. Due to the orthogonality of the transmitted signal, it can be derived $\boldsymbol{S}_p\boldsymbol{S}_{p^\prime}^H=\left\{\begin{matrix}\boldsymbol{0}_{K_t},\ p\neq p^\prime\\\boldsymbol{1}_{K_t},\ p=p^\prime\\\end{matrix}\right.$. Therefore, the output signal of the $p$-th matched filter of the $q$-th Rx RHS subarray, i.e., $\boldsymbol{Y}_{pq}=\boldsymbol{Y}_q\boldsymbol{S}_p^H$, can be expressed as\cite{5}:

\begin{equation}
\boldsymbol{Y}_{pq}=\left[\boldsymbol{\mathit{\Psi}}_q^r\left(\boldsymbol{Q}_q^r\circ\boldsymbol{\Gamma}_q^r\right)\right]^T\boldsymbol{V}_{pq},
\end{equation}
where
\begin{equation}
\boldsymbol{V}_{pq}=\sum_{l=1}^{L}\left(\beta_{pq}^l\boldsymbol{A}_{pq}^l\boldsymbol{X}_p\boldsymbol{J}_{pq}^l\right)+\boldsymbol{N}_{pq},
\end{equation}
where $\beta_{pq}^l$ is the reflection coefficient of the $l$-th target/clutter relative to the $(p, q)$-th transceiver array pair, and it is modeled as a Gaussian random variable with zero mean and variance ${\sigma_{pq}^l}^2$\cite{13}. Further, $\boldsymbol{A}_{pq}^l=\boldsymbol{a}_{q}^l\left(\vartheta_{q}^l,\varphi_{q}^l\right)\left(\boldsymbol{a}_{p}^l\left(\vartheta_{p}^l,\varphi_{p}^l\right)\right)^T$, where $\boldsymbol{a}_{q}^l\left(\vartheta_{q}^l,\varphi_{q}^l\right)\in\mathbb{C}^{N_r\times1}$ and $\boldsymbol{a}_{p}^l\left(\vartheta_{p}^l,\varphi_{p}^l\right)\in\mathbb{C}^{N_t\times1}$ respectively denote the steering vector of the $l$-th target/clutter with respect to the $q$-th Rx RHS and the $p$-th Tx RHS, with $\vartheta_{q}^l$, $\vartheta_{p}^l$, $\varphi_{q}^l$, and $\varphi_{p}^l$ being the corresponding azimuth and elevation angles. Besides, 
$\boldsymbol{J}_{pq}^l\in\mathbb{C}^{I_t\times I_r}$ is a shift matrix, representing the signal delay\cite{17}, and $\boldsymbol{N}_{pq}\in\mathbb{C}^{N_r\times I_r}$ is the matrix of the complex Gaussian noise of the $(p, q)$-th transceiver array pair after the matched filter, and the noise of the signal received by the $q$-th Rx RHS has zero mean and the variance of $\sigma_{n,q}^2=\sigma_n^2$. 

\section{Problem Formulation}
Given the false alarm probability, the target detection probability increases monotonically with signal-to-interference-plus-noise-ratio (SINR) under Gaussian noise conditions \cite{17}, and thus we use SINR as the research metric. The location information of targets is known as prior information\cite{aubry2013knowledge}. Considering clutters and other targets as interferences, the SINR of the echo signal reflected by the $l_t$-th target through the $(p,q)$-th transceiver array pair can be expressed as: 
\begin{equation}
{SINR}_{pq}^{l_t}\left(\boldsymbol{\psi}_p^t,\boldsymbol{\psi}_q^r\right)=\frac{\left|\boldsymbol{v}_{pq}^{l_t}(\boldsymbol{\psi}_p^t,\boldsymbol{\psi}_q^r)\right|^2}{\sum\limits_{l=1,l\neq l_t}^{L}\left|\boldsymbol{v}_{pq}^{l}(\boldsymbol{\psi}_p^t,\boldsymbol{\psi}_q^r)\right|^2+\left|\boldsymbol{w}_{pq}\right|^2},
\end{equation}
where $\boldsymbol{v}_{pq}^l(\boldsymbol{\psi}_p^t,\boldsymbol{\psi}_q^r)=vec\left[\beta_{pq}^l\left(\boldsymbol{\mathit{\Psi}}_q^r\left(\boldsymbol{Q}_q^r\circ\boldsymbol{\Gamma}_q^r\right)\right)^T\boldsymbol{A}_{pq}^l\boldsymbol{X}_p\boldsymbol{J}_{pq}^l\right]$
is the vectorized signal reflected by the $l$th target/clutter in $\boldsymbol{Y}_{pq}$, and $\boldsymbol{w}_{pq}=vec\left[\left(\boldsymbol{\mathit{\Psi}}_q^r\left(\boldsymbol{Q}_q^r\circ\boldsymbol{\Gamma}_q^r\right)\right)^T\boldsymbol{N}_{pq}\right]$ is the noise.

To balance the performance of all subsystems, we calculate ${\overline{SINR}}^{l_t}$ for each target $l_t$ as the average output SINR:
\begin{equation}
{\overline{SINR}}^{l_t}\left(\boldsymbol{\psi}^t,\boldsymbol{\psi}^r\right)=\sum_{p=1}^{P}\sum_{q=1}^{Q}{\frac{1}{PQ}{SINR}_{pq}^{l_t}\left(\boldsymbol{\psi}_p^t,\boldsymbol{\psi}_q^r\right)},
\end{equation}
where $\boldsymbol{\psi}^t=\left[\begin{matrix}{\boldsymbol{\psi}_1^t},...,{\boldsymbol{\psi}_P^t}\\\end{matrix}\right]^T\in\mathbb{R}^{{\rm PN}_t\times1}$ is defined for jointly designing all beamformers with 
$\boldsymbol{\psi}_p^t=\left(\boldsymbol{X}_p^t\boldsymbol{\psi}^t\right)^T$. Here $\boldsymbol{X}_p^t$ is a block matrix where the $p$-th block is $\boldsymbol{I}_{N_t}$ and the others are $\boldsymbol{0}_{N_t}$. Similarly, we have $\boldsymbol{\psi}^r=\left[\begin{matrix}{\boldsymbol{\psi}_1^r},...,{\boldsymbol{\psi}_Q^r}\\\end{matrix}\right]^T$ where $\boldsymbol{\psi}_q^r=\left(\boldsymbol{X}_q^r\boldsymbol{\psi}^r\right)^T$.

We aim to maximize the worst-case ${\overline{SINR}}^{l_t}$ over all targets\cite{17}, formulating the optimization problem as:
\begin{subequations} \label{eq:1}
\begin{align}
\text { P1: }\max _{\boldsymbol{\psi}^t, \boldsymbol{\psi}^r} \min _{l_t}&\left\{{\overline{SINR}}^{l_t}\left(\boldsymbol{\psi}^t, \boldsymbol{\psi}^r\right)\right\},\label{eq:2A}\\
s.t.\ \ &tr\left\{\left(\boldsymbol{\psi}^t\right)^H\boldsymbol{C}_p\boldsymbol{\psi}^t\right\}\le P_M,\ \forall p, \label{eq:2B}\\
& 0\le\psi_{{p,n}_t}^t\le1,\ \forall p,\forall n_t, \label{eq:2C}\\
& 0\le\psi_{{q,n}_r}^r\le1,\ \forall q,\forall n_r, \label{eq:2D}
\end{align}
\end{subequations}
where $\left(7b\right)$ indicates that the radiation power of each Tx RHS does not exceed the upper bound $P_M$. The matrix 
$\boldsymbol{C}_p=\left(\boldsymbol{X}_p^t\right)^H\left[\left(\left(\boldsymbol{Q}_p^t\circ\boldsymbol{\Gamma}_p^t\right)\boldsymbol{S}_p\right)\left(\left(\boldsymbol{Q}_p^t\circ\boldsymbol{\Gamma}_p^t\right)\boldsymbol{S}_p\right)^H\right]\circ\boldsymbol{I}_{N_t}\boldsymbol{X}_p^t$
is Hermitian positive semidefinite, so (7b) is a quadratic convex constraint. Constraints $\left(7c\right)$ and $\left(7d\right)$ represent the magnitude constraints of each RHS element, which are linearly convex. It is easy to prove that the intersection of all constraints in (P1) is not empty, so that (P1) is feasible.

In the optimization problem P1, $\boldsymbol{\psi}^t$ and $\boldsymbol{\psi}^r$ are coupled together, which is difficult to solve directly, so we decouple problem (P1) into two sub-problems and solve them iteratively.

    First, given beamformers for all Rx subarrays, namely $\boldsymbol{\psi}^r$, optimize beamformers for all Tx subarrays, namely $\boldsymbol{\psi}^t$.
    \begin{equation}
P2: \max _{\boldsymbol{\psi}^t} \ \min _{l_t}\left\{{\overline{SINR}}^{l_t}\left(\boldsymbol{\psi}^t\right)\right\}, \quad \text { s.t. }(7 b),(7 c).
    \end{equation}
    
    Second, given the beamformers of all Tx subarrays, i.e. $\boldsymbol{\psi}^t$, optimize the beamformers of all Rx subarrays, i.e. $\boldsymbol{\psi}^r$.
    \begin{equation}
P3: \max _{\boldsymbol{\psi}^r} \ \min _{l_t}\left\{{\overline{SINR}}^{l_t}\left(\boldsymbol{\psi}^r\right)\right\}, \text { s.t. }(7 d).
    \end{equation}

\renewcommand{\algorithmicrequire}{ \textbf{  Input:}}     
\renewcommand{\algorithmicensure}{ \textbf{ Output: }} 

\begin{algorithm}[t] 
\caption{Distributed RHS Radar Amplitude Optimization Algorithm (DRAOA)} 
\label{SA} 
\begin{algorithmic}[1] 
\REQUIRE Initialize ${\boldsymbol{\mathit{\Psi}}^t}^{(0)}=\boldsymbol{0}_{N_t+1}$, randomly initialize ${\boldsymbol{\mathit{\Psi}}^r}^{(0)}$;
\ENSURE $\boldsymbol{\psi}^{t\ }$, $\boldsymbol{\psi}^r$ and the objective function value ${\overline{SINR}}^{l_t}$;
\STATE Set $m=1$;
\REPEAT
\STATE Given ${\boldsymbol{\mathit{\Psi}}^r}^{(m-1)}$, solve (P2) to get ${\boldsymbol{\mathit{\Psi}}^t}^{(m)}$ and ${U^t}^{(m)}$;
\STATE Given ${\boldsymbol{\mathit{\Psi}}^t}^{(m)}$, solve (P3) to get ${\boldsymbol{\mathit{\Psi}}^r}^{(m)}$ and ${U^r}^{(m)}$;
\STATE Set $m=m+1$;
\UNTIL {$\left|{U^r}^{(m)}-{U^t}^{(m)}\right|\le\epsilon$};
\STATE Randomly generate $\boldsymbol{\zeta}_g^t\sim N\left(0,\boldsymbol{\mathit{\Psi}}^t\right),\ g=1,2\ ,...,\ G$, and $\boldsymbol{\zeta}_h^r\sim N\left(0,\boldsymbol{\mathit{\Psi}}^r\right),\ h=1,2,...,H$.
\STATE Replace the out-of-range elements in $\boldsymbol{\zeta}_g^t$ and $\boldsymbol{\zeta}_h^r$ to satisfy (7c) and (7d) and delete $\boldsymbol{\zeta}_g^t$ that does not satisfy (7b).
\STATE Select $\boldsymbol{\zeta}_g^t$ and $\boldsymbol{\zeta}_h^r$ that maximize the smallest ${\overline{SINR}}^{l_t}$ as $\boldsymbol{\psi}^{t\ }$ and $\boldsymbol{\psi}^r$, respectively.
\end{algorithmic}
\end{algorithm}

\section{Algorithm Design}
In this section, we propose a distributed RHS radar amplitude optimization algorithm (DRAOA) as summarized in Algorithm 1, which solves (P1) by iteratively solving (P2) and (P3), and then obtains $\boldsymbol{\psi}^{t\ }$ and $\boldsymbol{\psi}^r$ using Gaussian randomization (step7,8,9)\cite{10}. The algorithms for solving subproblems (P2) and (P3) are described in the following sections.

\subsection{Transmit Amplitude Optimization}
In this subsection, we aim to optimize the transmitting beamformer $\boldsymbol{\psi}^t$ given the receiving beamformer $\boldsymbol{\psi}^r$. Since (P2) is a max-min fractional sum problem, this is a non-convex NP-hard optimization problem. To facilitate the solution, firstly, we use positive semi-definite programming (SDP)\cite{10}. Define $\boldsymbol{\mathit{\Psi}}^t=\left[\begin{matrix}\boldsymbol{\psi}^t\\t\\\end{matrix}\right]\left[\begin{matrix}\boldsymbol{\psi}^t\\t\\\end{matrix}\right]^H$, where $t$ is a variable to homogenize SDP problem with $t^2=1$. Secondly, according to slack variable replacement (SVR)\cite{17}, we introduce two replacement variables $U^t=\min _{l_t} {\left\{{\overline{SINR}}^{l_t}\left(\boldsymbol{\mathit{\Psi}}^t\right)\right\}}$ 
and $\boldsymbol{\mathit\Lambda}=\left\{\lambda_{pq}^{l_t}\right\}_{p=1,q=1,{l_t}=1}^{P,Q,L_t}$, with $\lambda_{pq}^{l_t}={SINR}_{pq}^{l_t}\left(\boldsymbol{\mathit{\Psi}}^t\right)$. After relaxing the rank-one constraint of $\boldsymbol{\mathit{\Psi}}^t$, we express (P2) as 
\begin{subequations}\label{eq:3}
\begin{align}
P4: &\max _{\boldsymbol{\mathit{\Psi}}^t, U^t, \boldsymbol{\mathit\Lambda}} \ \ U^t,\label{eq:5A}\\
s.t.\ \ &Tr\left({\boldsymbol{C}_{p}^{'}}\boldsymbol{\mathit{\Psi}}^t\right)\le P_M, \ \forall p,\label{eq:5B}\\
&0\le Tr\left(\boldsymbol{X}_{{p,n}_t}\boldsymbol{\mathit{\Psi}}^t\right)\le1, \ \forall{p,n}_t,\label{eq:5C}\\
&\boldsymbol{\mathit{\Psi}}^t\succeq 0,\label{eq:5D}\\
&Tr\left(\boldsymbol{D}^t\boldsymbol{\mathit{\Psi}}^t\right)=1,\label{eq:5E}\\
&\sum_{p=1}^{P}\sum_{q=1}^{Q}{\frac{1}{PQ}\lambda_{pq}^{l_t}}\geq U^t,  \forall {l_t},\label{eq:5F}\\
\begin{split}
    &Tr\left[\left(\boldsymbol{R}_{pq}^{l_t'}-\lambda_{pq}^{l_t}\boldsymbol{R}_{pq}^{I'}\right)\boldsymbol{\mathit{\Psi}}^t\right] \geq Tr\left(\boldsymbol{R}_{pq}^{N'}\boldsymbol{\mathit{\Psi}}^r\right)\\&\lambda_{pq}^{l_t},\forall p, q, {l_t}.\\
\end{split}
\end{align}
\end{subequations}
where $\boldsymbol{R}_{pq}^{l_t'}$, $\boldsymbol{R}_{pq}^{I'}$, and $\boldsymbol{R}_{pq}^{N'}$ represent the matrix $\boldsymbol{R}_{pq}^{l_t}$, $\boldsymbol{R}_{pq}^{I}$, and $\boldsymbol{R}_{pq}^{N}$ expanded by one row and one column of zeros, respectively, and $\boldsymbol{R}_{pq}^{l_t}=\sum_{i_r=1}^{I_r}\left[\left(\boldsymbol{H}_{pq}^{l_t}\boldsymbol{X}_p^t\right)^H\left(\boldsymbol{H}_{pq}^{l_t}\boldsymbol{X}_p^t\right)\right]$ with $\boldsymbol{H}_{pq}^{l_t}={\beta}_{pq}^{l_t}\left(\boldsymbol{\mathit{\Psi}}_q^r\left(\boldsymbol{Q}_q^r\circ\boldsymbol{\Gamma}_q^r\right)\right)^T\boldsymbol{A}_{pq}^{l_t}diag\left(\boldsymbol{Q}_p^t\circ\boldsymbol{\Gamma}_p^t\boldsymbol{S}_p\boldsymbol{J}_{pq}^{l_t}(i_r)\right)$, $\boldsymbol{R}_{pq}^{I}=\sum_{l=1,l\neq {l_t}}^{L}\boldsymbol{R}_{pq}^{l}$, and $\boldsymbol{R}_{pq}^{N}=\sum_{i_r=1}^{I_r}\left(\boldsymbol{H}_{pq}^{N}\boldsymbol{X}_q^r\right)^H\left(\boldsymbol{H}_{pq}^{N}\right.\\\left.\boldsymbol{X}_q^r\right)$ with $\boldsymbol{H}_{pq}^{N}=\left(\boldsymbol{Q}_q^r\circ\boldsymbol{\Gamma}_q^r\right)^Tdiag\left(\boldsymbol{N}_{pq}(i_r)\right)$. The constraint $\left(10b\right)$ corresponds to the radiation power constraint $\left(7b\right)$, with ${\boldsymbol{C}_{p}}'$ representing the matrix $\boldsymbol{C}_{p}$ expanded by one row and one column of zeros. The constraint $\left(10c\right)$ corresponds to $\left(7c\right)$, which is a magnitude constraint, with $\boldsymbol{X}_{p,n_t}=\left[\begin{matrix}\boldsymbol{0}_{{PN}_t}&\boldsymbol{x}_{p,n_t}\\\left(\boldsymbol{x}_{p,n_t}\right)^T&0\\\end{matrix}\right]$, where $\boldsymbol{x}_{p,n_t}\in\mathbb{R}^{{\rm PN}_t\times1}$ represents a column vector whose $\left[\left(p-1\right)N_t+n_t\right]$-th element is 1/2 and the remaining elements are 0. 
The constraint $\left(10e\right)$ represents $t^2=1$, where $\boldsymbol{D}^t$ is a square matrix of dimensions $P{N_t}+1$ with all zeros except for the last element, which is 1.

At this point, (P4) is solved iteratively until achieving its convergence. The $k$-th iteration of solving (P4) consists of the following three steps.

First, since the objective and constraints in (P5) are convex, we solve (P5) using the CVX to obtain ${\boldsymbol{\mathit{\Psi}}^t}^{(k)}$, given ${U^t}^{(k-1)}$ and ${\boldsymbol{\mathit\Lambda}}^{(k-1)}$ which are computed in the $(k-1)$-th iteration.
\begin{equation}P5: \max _{\Psi^t} \ \ U^t, \quad \text { s.t. }(10 b)(10 c)(10 d)(10 e)(10 g).\end{equation}

Second, given ${\boldsymbol{\mathit{\Psi}}^t}^{(k)}$ from the first step, calculate ${\boldsymbol{\mathit\Lambda}}^{(k)}$ as 
    \begin{equation}
{\lambda_{pq}^{l_t}}^{(k)}=\frac{Tr\left(\boldsymbol{R}_{pq}^{l_t'}{\boldsymbol{\mathit{\Psi}}^t}^{(k)}\right)}{Tr\left(\boldsymbol{R}_{pq}^{I'}{\boldsymbol{\mathit{\Psi}}^t}^{(k)}\right)+Tr\left(\boldsymbol{R}_{pq}^{N'}\boldsymbol{\mathit{\Psi}}^r\right)}. 
    \end{equation}

Third, given ${\boldsymbol{\mathit\Lambda}^t}^{(k)}$ from the second step, calculate ${U^t}^{(k)}$ as ${U^t}^{(k)}=\min _{l_t} {\left\{\sum_{p=1}^{P}\sum_{q=1}^{Q}{\frac{1}{PQ}{\lambda_{pq}^{l_t}}^{(k)}}\right\}}$.

\subsection{Receive Amplitude Optimization}
Similar to solving the subproblem (P2), we first simplify the problem by using semi-definite relaxation (SDR) defining $\boldsymbol{\mathit{\Psi}}^r=\left[\begin{matrix}\boldsymbol{\psi}^r\\s\\\end{matrix}\right]\left[\begin{matrix}\boldsymbol{\psi}^r\\s\\\end{matrix}\right]^H$, with ${s}^2=1$. Then, we introduce two replacement variables: $U^r=\min _{l_t} {\left\{{\overline{SINR}}^{l_t}\left(\boldsymbol{\mathit{\Psi}}^r\right)\right\}}$, and $\boldsymbol{\mathit\Xi}=\left\{\xi_{pq}^{l_t}\right\}_{p=1,q=1,{l_t}=1}^{P,Q,L_t}$, where $\xi_{pq}^{l_t}={SINR}_{pq}^{l_t}\left(\boldsymbol{\mathit{\Psi}}^r\right)$, thus converting the optimization problem (P3) to the optimization problem (P6) as
\begin{subequations}\label{eq:20}
\begin{align}
P6: &\max _{\boldsymbol{\mathit{\Psi}}^{r}, U^{r},\boldsymbol{\mathit\Xi}} \ \ U^r,\label{eq:7A}\\
s.t.\ \ &0\le Tr\left(\boldsymbol{X}_{{q,n}_r}\boldsymbol{\mathit{\Psi}}^r\right)\le1,\forall q, n_r,\label{eq:20B}\\
&\boldsymbol{\mathit{\Psi}}^r\succeq 0,\label{eq:20C}\\ 
&Tr\left(\boldsymbol{D}^r\boldsymbol{\mathit{\Psi}}^r\right)=1,\label{eq:20D}\\
&\sum_{p=1}^{P}\sum_{q=1}^{Q}{\frac{1}{PQ}\xi_{pq}^{l_t}}\geq U^r, \forall {l_t},\label{eq:20E}\\
\begin{split}&\xi_{pq}^{l_t}Tr\left(\boldsymbol{M}_{pq}^{IN'}\boldsymbol{\mathit{\Psi}}^r\right)\le Tr\left(\boldsymbol{M}_{pq}^{l_t'}\boldsymbol{\mathit{\Psi}}^r\right),\forall p, q, {l_t},\end{split}
\label{eq:20F}
\end{align}
\end{subequations}
where $\left(13b\right)$ is the magnitude constraint, $\left(13c\right)$ is the constraint after positive semidefinite relaxation (SDR), and $\left(13d\right)$ represents ${s}^2=1$, where $\boldsymbol{D}^r$ is a square matrix of dimensions $Q{N_r}+1$ with all zeros except for the last element, which is 1. The constraints $\left(13e\right)$ and $\left(13f\right)$ are transformed from two substitute variables, where $\boldsymbol{M}_{pq}^{IN'}$ and $\boldsymbol{M}_{pq}^{l_t'}$ represent the matrix $\boldsymbol{M}_{pq}^{IN}$ and $\boldsymbol{M}_{pq}^{l_t}$ expanded by one row and one column of zeros, respectively, and $\boldsymbol{M}_{pq}^{l_t}=\sum_{i_r=1}^{I_r}{\left[\left(\boldsymbol{W}_{pq}^{l_t}\boldsymbol{X}_q^r\right)^H\boldsymbol{W}_{pq}^{l_t}\boldsymbol{X}_q^r\right]}$ with $\boldsymbol{W}_{pq}^{l_t}={\beta}_{pq}^{l_t}\left(\boldsymbol{Q}_q^r\circ\boldsymbol{\Gamma}_q^r\right)^Tdiag\left(\boldsymbol{A}_{pq}^{l_t}\boldsymbol{X}_p\boldsymbol{J}_{pq}^{l_t}(i_r)\right)$, and $\boldsymbol{M}_{pq}^{IN}=\sum_{l=1,l\neq {l_t}}^{L}\boldsymbol{M}_{pq}^{l}+\sum_{i_r=1}^{I_r}{\left(\boldsymbol{W}_{pq}^{N}\boldsymbol{X}_q^r\right)^H\boldsymbol{W}_{pq}^{N}\boldsymbol{X}_q^r}$ with $\boldsymbol{W}_{pq}^{N}=\left(\boldsymbol{Q}_q^r\circ\boldsymbol{\Gamma}_q^r\right)^T diag\left(\boldsymbol{N}_{pq}(i_r)\right)$. The method for solving problem (P6) is the same as problem (P5), which is omitted here.

\subsection{Complexity and Convergence Analysis}
The complexity of Algorithm 1 is mainly dominated by solving the SDP problems in (P4) and (P6), which are solved by the interior point method (IPM) with $\epsilon_{IPM}$ as the convergence parameter. Supposing that Algorithm 1, (P4), and (P6) are solved through $m$, $k^t$, and $k^r$ iterations respectively, the computational complexity\cite{10} of Algorithm 1 is 
$O\left(mk^t\sqrt{PQL_t+PN_t+P+2}\log\left(1/\epsilon_{\mathrm{IPM}}^{t}\right)(PN_t+1)^{3.5}\right)\\+O\left(mk^r\sqrt{PQL_t+QN_r+2}\log\left(1/\epsilon_{IPM}^{r}\right)(QN_r+1)^{3.5}\right)$.

In the $k$-th iteration when solving (P4), due to constraint (10f), ${U^t}^{(k)}\geq{U^t}^{(k-1)}$. In each iteration, $U^t$ is monotonically non-decreasing and has an upper bound due to the power constraint (10b), so the problem (P4) is convergent. Likewise, the problem (P6) is also convergent. When solving (P1) it is easy to introduce ${U^r}^{(m)}\geq{U^t}^{(m)}\geq{U^r}^{(m-1)}\geq{U^t}^{(m-1)}$ in the $m$-th iteration. Therefore, the smallest ${\overline{SINR}}^{l_t}$ is monotonically non-decreasing and has an upper bound due to the power constraint, so the Algorithm 1 is convergent.

\begin{figure*}[!t]
\centering
\subfloat[]{\includegraphics[width=2.36in]{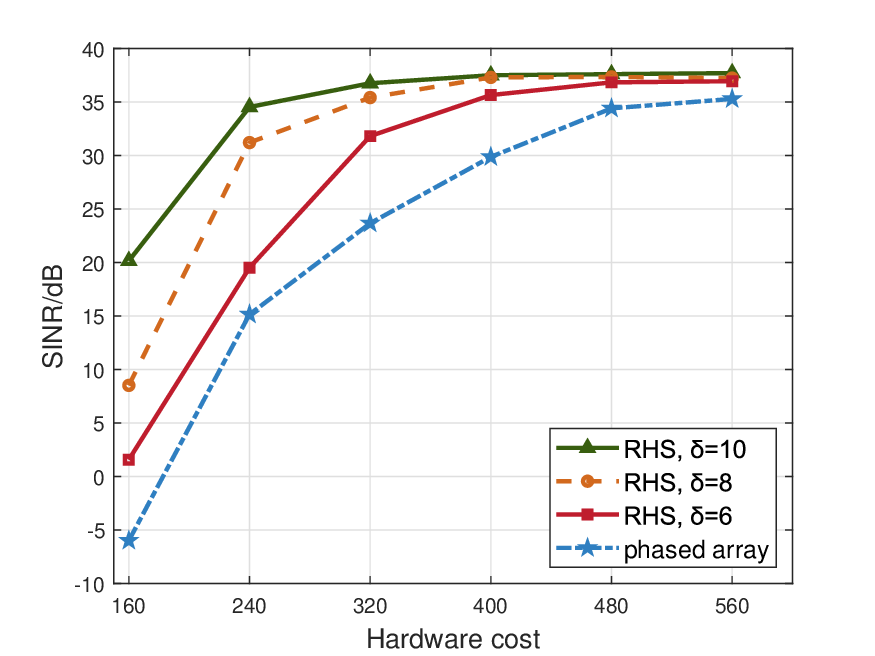}%
\label{fig_first_case}}
\hfil
\subfloat[]{\includegraphics[width=2.36in]{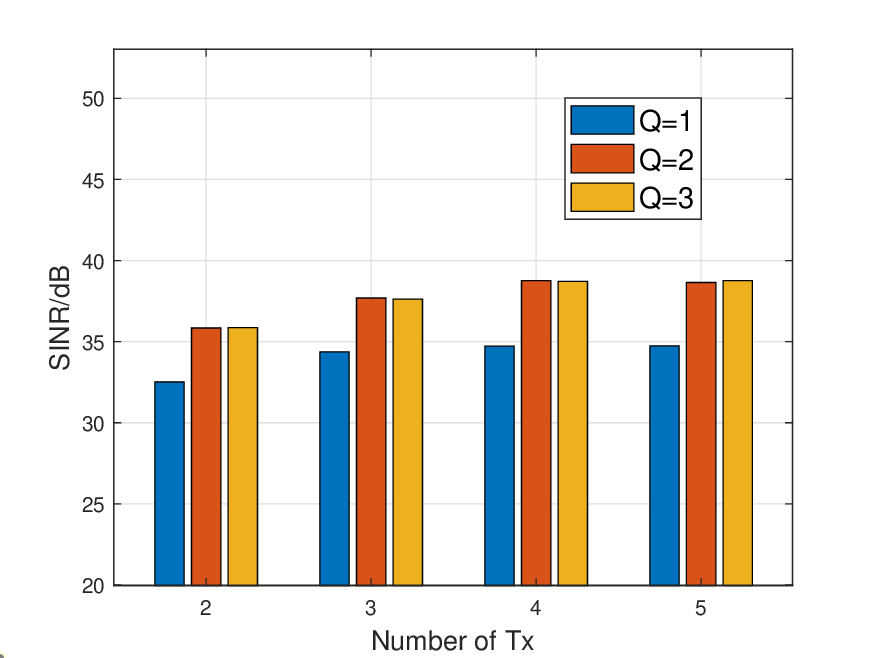}%
\label{fig_second_case}}
\hfil
\subfloat[]{\includegraphics[width=2.36in]{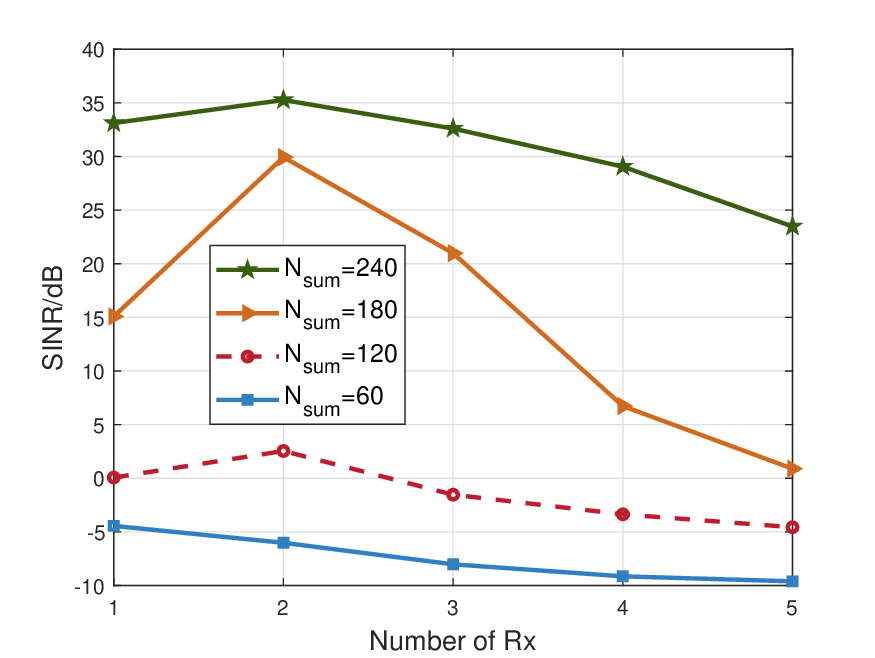}%
\label{fig_third_case}}
\caption{(a) SINR versus hardware cost; (b) SINR versus number of Tx; (c) SINR versus number of Rx.}
\label{fig_sim}
\end{figure*}

\section{Simulation Results}
In this section, simulation results are provided to evaluate the performance of the proposed scheme. The carrier frequency of the radar signal is $f_c=30GHz$, the wavelength $\lambda=1cm$, and the radiation power of each subarray $P_M=4mW$. The element spacing of RHS is $\lambda/3$, and the number of feeds per subarray is $K=5$. We set waveguide refraction coefficient $\nu=\sqrt3$ and amplitude attenuation factor $a=5$. Suppose that there are two targets, one clutter in the system, located at $\boldsymbol{p}_1=[1/2,\ 2\ ,1]$, $\boldsymbol{p}_2=[1,\ 3/2\ ,1]$, and $\boldsymbol{p}_3=[1,\ 2,\ 2]$, with power $\sigma_1^2=\sigma_2^2$, and $\sigma_3^2$, respectively. We set noise power ${\sigma}_n^2=4uW$, $SNR=10\log_{10}\left(\sigma_1^2/\sigma_n^2\right)=6dB$ and interference-to-noise ratio, i.e. $INR=10\log_{10}\left(\sigma_3^2/\sigma_n^2\right)=6dB$. We compare the proposed scheme with the distributed phased-MIMO scheme under the equivalent power consumption and hardware cost. The power consumption is defined as the total power consumed by the antenna, including the radiation power and the power consumed by the hardware circuits. According to the RHS prototype\cite{13}, the ratio of radiation power to total power of RHS and phased array is approximated as ${\eta}_R=25\%$ and ${\eta}_A=4\%$, respectively. On average, the hardware cost of a phased array antenna is $\delta=6\sim10$\cite{13} times that of an RHS element.

Fig. 2(a) shows SINR versus hardware cost of the proposed scheme (RHS) and the distributed phased-MIMO scheme (phased array) with $\delta=6, 8, 10$. For simplicity, the hardware cost of a phased array antenna is normalized to 10\cite{9}. Therefore, the hardware cost of an RHS element is $10/\delta$. Hardware cost denotes $10N_{sum}/\delta$. We set $P=Q=2$, and the hardware cost of the system is increased by increasing the number of elements per subarray. We observe that compared with the distributed phased-MIMO scheme under the equivalent power consumption and hardware cost, the SINR of the proposed RHS scheme exceeds at least 4.98dB on average, indicating that the proposed system achieves significantly better multi-target detection performance.

Fig. 2(b) depicts SINR versus number of Tx, i.e. $P$, with different number of Rx, i.e. $Q$. We set the number of elements per subarray $N_t=N_r=80$. It is shown that, given $Q$, SINR increases with the number of Tx and reaches saturation at $P=4$. Similarly, given $P$, SINR increases with the number of Rx and reaches saturation at $Q=2$. This indicates that increasing the number of Tx or the number of Rx brings spatial diversity gain, which improves the performance of multi-target detection until saturation.

Fig. 2(c) illustrates SINR versus number of Rx with different $N_{sum}$ of the system. The number of Tx $P$ is set to 2. It can be seen that when $N_{sum}=60$, SINR decreases as the number of Rx increases, which indicates that the spatial diversity gain brought by increasing Rx is not enough to compensate for the reduced signal coherence processing gain. When $N_{sum}=120, 180, 240$, SINR increases first and then decreases as the number of Rx increases, and reaches a peak at $Q=2$. It suggests that given the system hardware resources, there is an optimal system configuration, which achieves the optimal allocation between the signal coherent processing gain and the spatial diversity gain, so as to maximize the probability of multi-target detection.

\section{Conclusion}
In this letter, we have developed RHS-aided distributed MIMO radar systems for multi-target detection. In order to optimize the performance of the system, for each target, we have maximized the minimum SINR average of all radar subsystems, and have designed an optimization scheme for joint transceiver subarrays beamforming. Through the simulation results, we can draw the following conclusions: 1) Compared to distributed phased-MIMO radar systems with equivalent power consumption and hardware cost, the proposed scheme achieves superior multi-target detection performance.  2) Increasing the number of elements per subarray and the number of subarrays improves the performance of the system by enhancing the signal coherence processing gain and spatial diversity gain, respectively, and achieves saturation. 3) With fixed hardware resources, optimal system performance can be attained by appropriately allocating coherent processing gain and spatial diversity gain through the configuration of array elements. 
\bibliographystyle{IEEEtran}
\bibliography{ref,IEEEabrv}

\end{document}